\documentclass[%
reprint,
superscriptaddress,
footinbib,
longbibliography,
noeprint,
amsmath,amssymb,
aps,
prl,
]{revtex4-1}
\usepackage{amsthm}

\usepackage{xr}
\usepackage{graphicx}
\usepackage{physics}
\usepackage{xcolor}
\usepackage[autostyle, english = american]{csquotes}
\usepackage[ruled,vlined, linesnumbered]{algorithm2e}

\usepackage{bm}  
\renewcommand{\vec}[1]{\boldsymbol{\mathbf{#1}}}

\MakeOuterQuote{"}

\definecolor{darkGreen}{RGB}{0,110,0}
\definecolor{darkBlue}{RGB}{0,0,130}
\usepackage[colorlinks,citecolor=darkGreen,linkcolor=darkBlue,urlcolor=blue,hyperindex]{hyperref}

\def\equationautorefname~#1\null{Eq. (#1)\null}

\newcommand{\appref}[1]{\hyperref[#1]{App.~\ref*{#1}}}

\newcommand{\comment}[1]{}

\begin{document}
\date{\today} 

\title{Arresting Dynamics in Hardcore Spin Models}

\author{Benedikt Placke}
\affiliation{Max-Planck-Institut f\"{u}r Physik komplexer Systeme, 01187 Dresden, Germany}

\author{Grace M. Sommers}
\affiliation{Department of Physics, Princeton University, Princeton, NJ 08540, USA}

\author{S. L. Sondhi}
\affiliation{Department of Physics, Princeton University, Princeton, NJ 08540, USA}
\affiliation{Rudolf Peierls Centre for Theoretical Physics, University of Oxford, Oxford OX1 3PU, United Kingdom}

\author{Roderich Moessner}
\affiliation{Max-Planck-Institut f\"{u}r Physik komplexer Systeme, 01187 Dresden, Germany}

\begin{abstract}
We study the dynamics of hardcore spin models on the square and triangular lattice, constructed by analogy to hard spheres, where the translational degrees of freedom of the spheres are replaced by orientational degrees of freedom of spins on a lattice and the packing fraction as a control parameter is replaced by an exclusion angle. 
In equilibrium, models on both lattices exhibit a Kosterlitz-Thouless transition at an exclusion angle $\Delta_{\rm KT}$.
We devise compression protocols for hardcore spins and find that {\it any} protocol that changes the exclusion angle nonadiabatically, if endowed with only local dynamics, fails to compress random initial states beyond an angle $\Delta_{\rm J}> \Delta_{\rm KT}$.
This coincides with a doubly algebraic divergence of the relaxation time of compressed states towards equilibrium.
We identify a remarkably simple mechanism underpinning this divergent timescale: topological defects involved in the phase ordering kinetics of the system become incompatible with the hardcore spin constraint, leading to a vanishing defect mobility as $\Delta\rightarrow\Delta_{\rm J}$.
\end{abstract}

\maketitle

\paragraph{Introduction.}

Within the realm of condensed-matter physics, examples abound of continuum systems whose critical phenomena can be well captured by simplified lattice models. For instance, the liquid-gas transition can, remarkably, be described using the lattice gas model, which then maps onto the Ising model~\cite{Goldenfeld1992}. Similarly, the Edwards-Anderson model and related spin-glass Hamiltonians provide a fruitful avenue toward understanding random impurities in magnetic alloys~\cite{Edwards1975}.

In this letter, we take a lattice approach towards what might be called vitrifaction, i.e.\ the emergence of slow dynamics which has played an important role under the headings of glassiness, freezing, jamming and the like. To do so, we devise a particularly simple yet versatile and tractable family of models of what we have termed hardcore spins. These insulating (non-itinerant) spin models are constructed by analogy to hard spheres/disks, which are the common idealized model systems for the phenomenon of jamming. Much progress has been made to understand the random jamming transition of hard spheres both numerically \cite{OHern2003, Berthier2011} as well as in infinite dimensions \cite{Parisi2010}. In particular, it is possible to define random jamming without the need to invoke any particular dynamics, the hallmark being a jump of the contact number $z$ as a function of packing fraction at “point J” $\Phi_J$. 

In contrast, our work takes a different tack: addressing the broad issue of dynamical arrest of an athermal system---one in which thermal fluctuations make a negligible contribution to its dynamics---far from equilibrium, we study whether, and by what mechanism, this phenomenon can arise in a lattice model. Concretely, we address its relation to the physics of phase ordering kinetics, i.e.\ the extent to which an underlying, but under a given dynamics inaccessible, ordered state and its concomitant topological defects play a fundamental role, an idea that dates back several decades in the glass literature \cite{Halperin1978, Nelson1979, Nelson1983, Nussinov2018, Scalliet2019}.


\begin{figure}
	\includegraphics{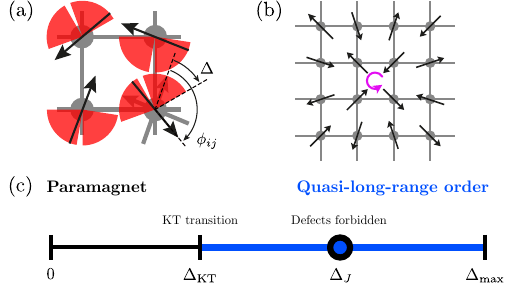}
	\caption{(a) Illustration of the hardcore constraint. For a given spin, red shading indicates the orientations forbidden by its nearest neighbors. (b) An antiferromagnetic vortex on the square lattice. The vortex core is marked in purple. (c) Minimal phase diagram of the model.}
	\label{fig:model}
\end{figure}

The hardcore spin models are defined by a local constraint, whereby no two neighboring hardcore spins are allowed to enclose an angle smaller than their exclusion angle [see \autoref{fig:model} (a)]. This can be viewed as an orientational, lattice analog to the non-local constraint on the translational degrees of freedom of hard spheres, for which two sphere centers must not be closer than the sum of their radii. The exclusion angle $\Delta$ serves as a tuning parameter for the equilibrium behavior of the system, which we have investigated in Ref. \onlinecite{Sommers2020}. The simplest possible phase diagram of hardcore spins is shown in \autoref{fig:model} (c). At low exclusion angle, the system is weakly constrained and hence in a paramagnetic phase. As the exclusion angle is increased in equilibrium, the system undergoes an entropically driven Kosterlitz-Thouless (KT) transition at $\Delta_{\rm KT}$.

Crucially, in a constrained system topological defects can not only become confined but their existence can even become strictly incompatible with the hardcore constraint. This happens at a point $\Delta_{\rm J} > \Delta_{\rm KT}$, and the defect density vanishes with a power law as $\Delta \to \Delta_{\rm J}$\cite{supplements}.

In this letter, we generalize compression protocols originally developed for hard spheres~\cite{Lubachevsky1990} and study dynamics of hardcore spins far from equilibrium.
Importantly, we uncover a remarkably simple mechanism precipitating of what might be called ``arresting'' or perhaps ``jamming dynamics'' in our model. In particular, any adiabatic compression protocol, namely one in which the system stays in equilibrium as $\Delta$ is driven through $\Delta_{\rm J}$, reaches the ordered state with maximal exclusion angle $\Delta_{\rm max}$, analogous to the close packing. In contrast, a nonadiabatic protocol starting from a random initial state at $\Delta=0$ will fall out of equilibrium and reach $\Delta = \Delta_{\rm J}$ with nonzero defect density. As the existence of defects is incompatible with the hardcore constraint for $\Delta > \Delta_{\rm J}$, the protocol will fail beyond this point.
The failure of compression at $\Delta_{\rm J}$ coincides with a doubly algebraic divergence of the relaxation time of compressed states towards equilibrium---as a function of both system size as well as the distance to $\Delta_{\rm J}$. This can be fully understood in terms of a diffusion-annihilation process of topological defects, with a vanishing defect mobility.

In the remainder of this letter we first present hardcore XY spins on the square lattice as a minimal model exhibiting all the abovementioned features. 
We then show that on the triangular lattice, the relevant physics is realized in a more complex way: the additional chiral symmetry in the model leads to the presence of two kind of topological defects, that is, domain walls in addition to vortices. This leads to a more structured slow dynamics, but crucially, the behavior of the model can be accounted for by the same fundamental mechanism as before.

\paragraph{Model and phase diagram on the square lattice.}
We consider a system of XY spins, that is two-component unit vectors $\{\vec S_j\}$, on a lattice $\mathcal L$ subject to the constraint
\begin{equation}
	\phi_{ij} := \arccos(\vec S_i \cdot \vec S_j) > \Delta~~\forall \expval{ij} \in \mathcal L,
	\label{eq:hardcore-spins}
\end{equation}
where $\expval{ij}$ denotes an edge between site $i$ and $j$.

For XY spins on the square lattice, the hardcore constraint is illustrated in \autoref{eq:hardcore-spins} (a), where for each spin, we denote in red the orientations forbidden while keeping all neighbors fixed. 
This model realizes the minimal equilibrium phase diagram sketched in \autoref{fig:model} (a). At $\Delta = 0$, the system is unconstrained and naturally in a paramagnetic state, while at $\Delta = \pi$, the only allowed state (up to global symmetry) is the N\'eel state. 
Between these extremes, for any $\Delta < \Delta_{\rm max}$, long-ranged order is prohibited by an extension of the Mermin-Wagner theorem to constrained systems \cite{Mio2015,Peled2021}.
The system instead undergoes a KT transition at $\Delta_{\rm KT}\approx0.435\pi$ \cite{Bietenholz2013a}, into a phase with quasi-long-ranged (QLR) order, with algebraically decaying correlations. 
Note that the model has no inherent notion of energy or temperature. Instead, \autoref{eq:hardcore-spins} separates states into two classes -- allowed and disallowed -- without endowing them either with dynamics or with a notion of (high and low) energy. Thus, nontrivial correlations are of entropic origin, i.e. a form of “order by disorder” (OBD) familiar from the nematic ordering of Onsager’s hard rods \cite{Onsager1949} and its descendants, including a recurrent appearance in frustrated magnetic systems.

It is well known that the KT transition can be understood as an unbinding transition of topological defects, which on the square lattice are vortices, see \autoref{fig:model} (b). Defect unbinding is indeed the mechanism behind the equilibrium transition in our model, but within the QLRO phase, these vortices becomes incompatible with \autoref{eq:hardcore-spins} for $\Delta > \pi/2$. This is shown most easily by realizing that on a bipartite graph such as the square lattice, any exclusion model as defined by \autoref{eq:hardcore-spins} maps on an inclusion model defined by 
\begin{equation}
	\tilde \phi_{ij} < \Delta_{\rm incl}~~\forall \expval{ij} \in \mathcal L,
	\label{eq:inclusion-model}
\end{equation}
where the spins on one of the two sublattices are flipped
\begin{equation}
    \tilde{\vec S}_j = \begin{cases}
        \phantom{-}\vec S_j & \text{if $j$ in sublattice $A$}\\
        -\vec S_j & \text{if $j$ in sublattice $B$},
    \end{cases}
\end{equation}
and $\Delta_{\rm incl} = \pi - \Delta$.
An antiferromagnetic vortex then maps onto a regular ferromagnetic vortex. Such a vortex however must have a core \cite{Sommers2020}, which is a single plaquette with a winding of $2\pi$. Since there are only four sites around a plaquette, this is only possible if $\Delta_{\rm incl} > 2\pi/4 = \pi/2$, which implies $\Delta < \pi/2$ in the exclusion model. 

\paragraph{Jamming hardcore spins.}

As a first step towards studying our model far from equilibrium, we define a notion of compression, which here will mean an increase of the exclusion angle $\Delta$ while simultaneously evolving the state under some local dynamics. 
An analogous protocol for hard disks was implemented by Lubachesvky and Stillinger \cite{Lubachevsky1990}. In this work, the radii of the disks were increased during a molecular dynamics simulation, while keeping the volume fixed. When the compression rate was slow with respect to the time scale set by the molecular dynamics, the final state was approximately close packed. In contrast, when the radii were increased very fast, the system ended up ``jammed'' in a polycrystalline state (to be contrasted with ``maximally random jammed'' states \cite{Atkinson2014}), at a density well below close packing.

The Lubachevsky-Stillinger (LS) algorithm is straightforwardly adapted to and implemented for hardcore spins, the most subtle point being the choice of dynamics.
Arguably the closest analog to the molecular dynamics studied by LS is given by continuous-time "Hamiltonian" dynamics. For hardcore spins, however, such dynamics are not ergodic even at infinitesimal exclusion angles since they preserve the local vorticity on each plaquette~\cite{supplements} (see also references \cite{Jepsen1965, Percus1969} therein).
Because of this, in the main text, we focus on a different kind, that is discrete-time, stochastic Monte-Carlo dynamics. In each time step, a single Monte-Carlo move is performed. Each such move consists of choosing a random site in the lattice, then choosing a random new configuration for the spin on the site and accepting the move if and only if the new configuration is allowed by \autoref{eq:hardcore-spins}. A similar technique was previously applied in the compression of hard spheres in Ref.~\onlinecite{Berthier2009}.

The compression protocol then proceeds as follows. Starting from a random state at $\Delta = 0$, one alternates $N_{\rm rattle}$ Monte-Carlo sweeps (with one sweep consisting of $N$ Monte-Carlo moves) and an increase of the exclusion angle by $\delta$, where the increment $\delta$ is chosen as large as possible without invalidating the current state. Fast compression in this context then means large $\delta / N_{\rm rattle}$. The protocol terminates if the increment $\delta$ repeatedly falls below a threshold value. 

\begin{figure}
	\includegraphics{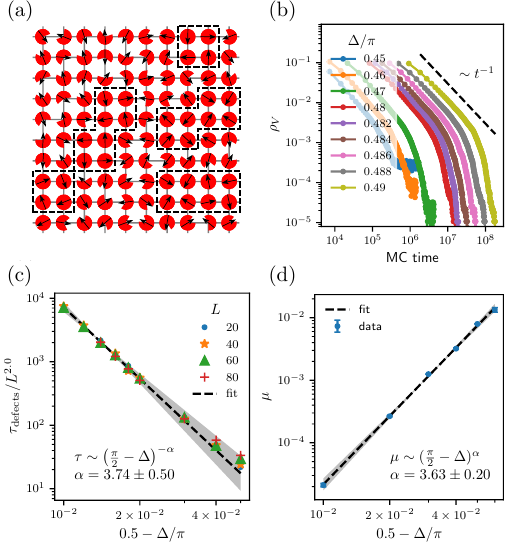}
	\caption{(a) State at $\Delta = \Delta_{\rm J}$, frozen vortex cores are enclosed by dashed lines. (b) Vortex  density $\rho_{\rm V}$ of far-from-equilibrium states relaxing towards equilibrium at different exclusion angles $\Delta$, for $L=60$. (c) Doubly algebraic behavior of relaxation time scale fitted from long-time tail of $\rho_{\rm V}$ as $\Delta \to \Delta_{\rm J}$. (d) Vanishing defect mobility as $\Delta \to \Delta_{\rm J}$. }
	\label{fig:square}
\end{figure}

On a $L = 40$ square lattice, we run the LS compression protocol for hardcore spins with different values of $N_{\rm rattle}$ to see whether jamming dynamics is observed.
Out of 100 runs, with $N_{\rm rattle} = 100$ all runs terminate at $\Delta_{\rm max} = \pi$, that is they reach the N\'eel state which is our analog to close packing. In contrast, for $N_{\rm rattle} = 1$ all runs terminate at $\Delta_{\rm J} = \pi/2$. For $N_{\rm rattle} = 10$, runs fall into two classes, with 16 terminating at $\Delta_{\rm J}$ and the rest at $\Delta_{\rm max}$. 
The ``jammed'' states at $\Delta_{\rm J}$ have a finite ordered moment, which increases with $N_{\rm rattle}$ but is always lower than the equilibrium expectation value \cite{supplements}.
To illustrate that the arresting dynamics here is indeed explained by our abovementioned mechanism, we show in \autoref{fig:square} (a) a state at $\Delta_{\rm J}$ for $L=10$ and indicate the frozen vortex cores by dashed black boxes. These vortex cores are completely frozen under local dynamics since the constituent spins enclose an angle of exactly $\pi/2$.

\paragraph{Relaxation dynamics.}
To corroborate the picture of defect freezing as the mechanism for the failure of nonadiabatic compression beyond $\Delta_{\rm J}$ in our model, we study relaxation dynamics towards equilibrium close to 
this point
in more detail. 
Generally speaking, one expects such a freezing to appear in conjunction with a diverging relaxation time scale.

While the LS protocol suffices to demonstrate the presence of jamming dynamics in hardcore spins, it is limited in that it allows only moderate compression speeds, resulting in partial equilibration, evidenced by a finite ordered moment of the final states.
Because of this, we implement a second kind of compression protocol, originally developed by Xu et al. \cite{Xu2005}, which utilizes a softened constraint to compress faster and hence avoid partial equilibration. 
This is done by introducing an energy functional
\begin{equation}
    V(\phi_{ij}, \Delta) =
    \begin{cases}
        \frac{1}{2}\left( 1 - \phi_{ij}/\Delta\right)^2 & \text{for }\phi_{ij} < \Delta \\
        0 & \text{for }\phi_{ij} \geq \Delta
    \end{cases},
    \label{eq:softcore-spins}
\end{equation}
which is zero if \autoref{eq:hardcore-spins} is fulfilled and introduces a quadratic energy penalty if neighboring hardcore spins overlap.
The introduction of the soft constraint enables us to use much larger increments $\delta$. This is because one does not have to choose $\delta$ such that the current state of the system stays valid, but instead one can choose it such that a conjugate gradient minimization step after the increment recovers a state with zero energy.

The softcore compression protocol consistently yields far-from-equilibrium states with an ordered moment close to zero. We prepare such states at a range of exclusion angles $\Delta$ close to $\Delta_{\rm J}$ and show their defect density as a function of Monte-Carlo time in \autoref{fig:square} (b). As expected from scaling arguments \cite{Yurke1993}, it decays diffusively initially, but at long times, this behavior gives way to an exponential decay with a characteristic relaxation time $\tau_{\rm defects}$. This relaxation time as a function of exclusion angle $\Delta$ is shown in \autoref{fig:square} (c), for a range of different system sizes. Evidently, the data is consistent with a doubly algebraic behavior
\begin{equation}
    \tau \sim L^z \left( \Delta_J - \Delta \right)^{-\alpha},
    \label{eq:relaxation-time}
\end{equation}
with $z=2$ and $\alpha = 3.74 \pm 0.50$.
These two power laws are conceptually quite distinct. The divergence of the relaxation times with system size, $\tau \sim L^z$ owes its existence to universal long-wavelength physics of phase-ordering kinetics under local dynamics and consequently its value $z=2$ \cite{Jensen2000} is quite robust. 
In contrast, the divergence of relaxation time as a function of exclusion angle is related to the defect mobility $\mu$. 
It is measured by preparing two isolated vortices in an otherwise paramagnetic state. Their distance then follows the time dependence $D(t) = \sqrt{D_0 - \mu t}$, from which $\mu$ can be determined by a linear fit. 
As shown in \autoref{fig:square} (d), it vanishes as a power law as $\Delta\rightarrow \Delta_J$ with, for a given local dynamics, roughly the same exponent $\alpha$ as the relaxation time. However, this exponent can be readily varied by changing the rules of the dynamical evolution even locally. For example, results from studying phase ordering kinetics under Hamiltonian dynamics plus tunneling are consistent with \autoref{eq:relaxation-time}, with $z=2$ but $\alpha=1.87\pm0.02$ \cite{supplements}.

\paragraph{Triangular lattice.}

\begin{figure}
    \centering
    \includegraphics{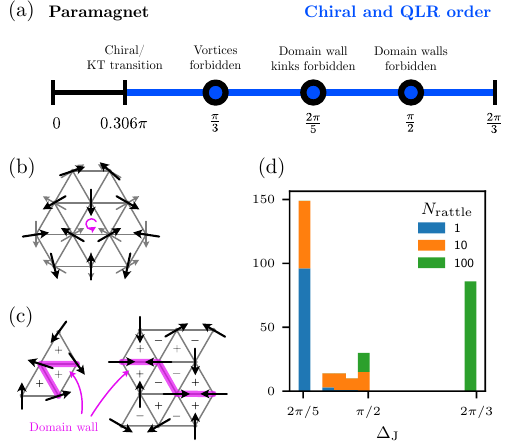}
    \caption{(a) Equilibrium phase diagram of the model on the triangular lattice.
    (b) A vortex on the triangular lattice.
    (c) Domain walls of chirality both with a kink (left) and without any kinks (right).
    (d) Distribution of jamming angles $\Delta_J$ as a function of compression speed for $L=42$; lower $N_{\rm rattle}$ implies faster compression.}
    \label{fig:triangular}
\end{figure}

Finally, we compare the simple picture of the square lattice model to that on the triangular lattice. Its phase diagram, \autoref{fig:triangular} (a), also has a paramagnetic phase including $\Delta=0$, with a single (up to global symmetries) three-sublattice ordered state at $\Delta=2\pi/3$; this comes in two chiralities that are not related by global rotation but instead by exchange of two sublattices.
In between, chiral and quasi-long-range order develop either simultaneously or at two separate transitions~\cite{Obuchi2012, Okumura2011} in proximity to $\Delta \approx 0.306\pi$ \cite{supplements} (see also references \cite{Lee1998, Hasenbusch2005b, Hasenbusch2008, Hasenbusch2005, Kamieniarz1993} therein).
 
The properties of topological defects, and in particular their preclusion upon increasing $\Delta$ remain central to the jamming phenomenology. 
These are richer than in the square lattice case. Vortices, as shown in \autoref{fig:triangular} (b), become forbidden at a single packing fraction $\Delta = \pi/3$. In contrast, domain walls have internal structure (their shape) and become incompatible with the hardcore constraint over a range of exclusion angles. First, kinks in domain walls, shown on the left of \autoref{fig:triangular} (c) become forbidden at $\Delta = 2\pi/5$ while at $\Delta = \pi/2$, {\it any} domain wall becomes incompatible with \autoref{eq:hardcore-spins} \cite{supplements}. 

In \autoref{fig:triangular} (d), we show a histogram of the angle at which compression of random initial states using the LS protocol at different $N_{\rm rattle}$ terminates. On the square lattice, such a histogram is strictly bimodal, with two narrow peaks at $\Delta_{\rm J} = \pi/2$ and $\Delta_{\rm max} = \pi$. In contrast, on the triangular lattice, for intermediate compression speed $N_{\rm rattle} = 10$, we see a continuous distribution between $\Delta = 2\pi/5$ with a pronounced peak at $\Delta=\pi/2$.
This can be understood by considering the same mechanism as on the square lattice: fast compression (that is small $N_{\rm rattle}$) fails because it arrives with a finite density of topological defects at a point where these become incompatible with \autoref{eq:hardcore-spins}. Now if there are multiple kinds of defects in the system, these can have different relaxation time scales and hence lead to a more complex dependence of the jamming angle on compression speed.

In this picture, it is somewhat surprising that in \autoref{fig:triangular} (d), there is no peak at $\Delta = \pi/3$, where vortices become forbidden. This is just a consequence of the fact that vortices are only well defined in the presence of chiral order, whereas we start from random initial states, that are neither chirally nor QLR ordered and have no well-defined vortex density to begin with. A random initial state will however have a well defined and large domain wall density, which leads to the failure of fast compression (jamming) at $\Delta_{\rm J} = 2\pi/5$, which is where kinked domain wall become forbidden. At intermediate speeds, the system then is able to partially equilibrate and reach zero domain-wall-kink density, but still has smooth domain walls, leading to jamming at $\Delta_{\rm J} = \pi/2$. Between these two points, in the range $2\pi/5 < \Delta < \pi/2$, there exist a multitude of local clusters with long but finite relaxation times, leading to jamming of protocols with intermediate compression speeds. 

An important, qualitative difference to the square lattice model is that on the triangular lattice the only \emph{vanishing} defect mobility in the model is that of domain walls at $\Delta = \pi/2$. Still, we observe a failure of the compression protocols at other points, where different defects become forbidden without a concomitant freezing. This is not unexpected since a long but finite relaxation time of such defects might still bring compression to a halt. In the language of granular materials, such freezing is \emph{fragile} in that rattling of a jammed state might unjam it. 

\paragraph{Conclusion.}
In summary, we have provided a detailed phenomenology and comprehensive understanding of arresting dynamics in hardcore spin models, uncovering an intricate interplay of lattice geometry, ordering and defects, and the dynamics at long and short wavelengths. Particularly noteworthy from a conceptual perspective is the role played by the (in)ability to anneal  defects under a purely local dynamics. This fundamentally accounts for the phenomenon in this \textit{lattice} model, with considerable added richness as a result of the variability and multiplicity of defects and their configurations. In addition to lending this model intrinsic interest on its own, the central role of defects to the (slow) dynamics resonates with a possibility previously formulated for the case of the glass transition \cite{Nussinov2018}. 

\begin{acknowledgments}
This work was in part supported by the Deutsche Forschungsgemeinschaft under grants SFB 1143 (project-id 247310070) and the cluster of excellence ct.qmat (EXC 2147, project-id 390858490). 
GMS is supported by the Department of Defense (DoD) through the National Defense Science \& Engineering Graduate (NDSEG) Fellowship Program and SLS would like to acknowledge the support of the Department  of  Energy  via  grant  No.   DE-SC0016244. Additional support was provided by the Gordon and Betty Moore Foundation through Grant GBMF8685 towards the Princeton theory program and the use of the TIGRESS High Performance Computer Center at Princeton University.
\end{acknowledgments}

\bibliography{references.bib}

\end{document}


\date{\today} 

\title{Supplementary Material to ``Arresting Dynamics in Hardcore Spin Models''}

\author{Benedikt Placke}
\affiliation{Max-Planck-Institut f\"{u}r Physik komplexer Systeme, 01187 Dresden, Germany}

\author{Grace M. Sommers}
\affiliation{Department of Physics, Princeton University, Princeton, NJ 08540, USA}

\author{S. L. Sondhi}
\affiliation{Department of Physics, Princeton University, Princeton, NJ 08540, USA}
\affiliation{Rudolf Peierls Centre for Theoretical Physics, University of Oxford, Oxford OX1 3PU, United Kingdom}

\author{Roderich Moessner}
\affiliation{Max-Planck-Institut f\"{u}r Physik komplexer Systeme, 01187 Dresden, Germany}

\maketitle

\tableofcontents

%

\section{Compression protocols for hardcore spins}\label{sect:protocols}

\subsection{Lubachevsky-Stillinger compression protocol}

\begin{algorithm}
\SetAlgoLined
\SetKwInOut{Output}{Output}
\SetKwInOut{Input}{Input}
\Input{Trial increment $\delta_{0}$\\
Minimum increment $\delta_{\rm min}$\\
Number of rattling sweeps $N_{\rm rattle}$}
\Output{Valid configuration $\{\vec{S}_i\}$ at $\Delta_J$,\\Final exclusion angle $\Delta_J$}
$\mathcal S \gets$ random initial configuration\;
$\Delta \gets 0$\;
\While{$\delta > \delta_{\rm min}$}{
$\delta \gets \delta_0$\;
\While{$\mathcal S$ \normalfont{is not valid at} $\Delta + \delta$}{
Perform $N_{\rm rattle}$ Monte-Carlo sweeps\;
$\delta \gets \delta/2$\;
}
$\Delta \gets \Delta + \delta$
}
\Return $\mathcal S$, $\Delta$\;

\caption{Lubachevsky-Stillinger compression\label{alg:LS-compression}}
\end{algorithm}

The Lubachevsky-Stillinger protocol, inspired by Ref. \onlinecite{Lubachevsky1990} and described in the main text, was implemented in \CC. It is given in pseudocode in \autoref{alg:LS-compression}. In \autoref{fig:hists}, we summarize the statistical properties of the final states. In particular, we show histograms of the final exclusion angle in the top row and histograms of the structure factor peak value in the bottom row. 

The fundamental limitation of this protocol is that what determines the largest possible increment of the inclusion angle $\Delta$ is the smallest angle between \emph{any} nearest neighbor pair in the system. This value however vanishes in the limit of infinite system size $N \to \infty$, implying that for fixed  $N_{\rm rattle}$ also the effective compression speed $\expval{\delta/N_{\rm rattle}} \to 0$ as $N \to \infty$. For $L=40$ we find $\expval{\delta}\approx 10^{-5}$.

\subsection{Softcore protocol}

\begin{algorithm}
\SetAlgoLined
\SetKwInOut{Output}{Output}
\SetKwInOut{Input}{Input}
\Input{Valid configuration $\mathcal S_0 = \{\vec{S}_i\}$ at $\Delta_0$,\\Trial increment $\delta_0$,\\Tolerance $\epsilon$}
\Output{Valid configuration $\{\vec{S}_i\}$ at $\Delta$,\\New exclusion angle $\Delta$}
$\mathcal S \gets \mathcal S_0$\;
$V \gets V^{(\rm tot)}(\mathcal S, \Delta + \delta)$\;
MODE $\gets$ COMPRESS\;
$\delta \gets \delta_0$\;
$\Delta \gets \Delta_0+ \delta$\;
\While{$V > \epsilon$ {\bf or} $V < \epsilon /2$} {
$\mathcal S_{\rm old} \gets \mathcal S$\;
$\mathcal S$, $V$, $\gets$ Conjugate gradient descent of $V^{(\rm tot)}(\mathcal S, \Delta)$ with initial state $\mathcal S$\;
\If{$V > \epsilon$ }{
\If{MODE $=$ COMPRESS}
{
$\delta \gets \delta/2$\;
MODE $\gets$ EXPAND\;
}
$\mathcal S \gets \mathcal S_{\rm old}$\;
$\Delta \gets \Delta - \delta$\;
}
\If{$V < \epsilon$}{
\If{MODE $=$ EXPAND}{
$\delta \gets \delta/2$\;
MODE $\gets$ COMPRESS\;
}
$\Delta \gets \Delta + \delta$\;
}
}
\Return $\mathcal S$, $\Delta$ \;
\caption{Softcore Compression Step \texttt{Compress}\label{alg:compression}}
\end{algorithm}

\begin{algorithm}
\SetAlgoLined
\SetKwInOut{Output}{Output}
\SetKwInOut{Input}{Input}
\Input{Max, min step size $\delta_{\rm max}$, $\delta_{\rm min}$\\
Expansion rattling sweeps $N_{\rm rattle}^{(\rm exp)}$\\
Extra rattling sweeps $N_{\rm rattle}$\\
Maximum number of expansions $N_{\rm exp}^{(\rm max)}$\\
Expansion step $\delta_{\rm exp}$}
\Output{Valid configuration $\mathcal S = \{\vec{S}_i\}$ at $\Delta$\\
Final exclusion angle $\Delta$}
$\Delta \gets 0$\;
$\mathcal S \gets$ random configuration\;
\While{$N_{\rm exp} < N_{\rm exp}^{(\rm max)}$}{
$N_{\rm exp} \gets 0$\;
$\Delta_{\rm max} \gets \Delta$\;
$\delta \gets 0$\;
\While{$\delta < \delta_{\rm min}$ {\bf and} $N_{\rm exp} < N_{\rm exp}^{(\rm max)}$}{
$\mathcal S, \Delta \gets$ Compress($\mathcal S, \delta_{\rm max}$)\;
$\delta \gets \Delta - \Delta_{\rm max}$\;
\If{$\delta < \delta_{\rm min}$}{
$N_{\rm exp} \gets N_{\rm exp} + 1$\;
$\Delta \gets \Delta - N_{\rm exp} \cdot \delta_{\rm exp}$\;
Perform $N_{\rm rattle}^{\rm (exp)}$ Monte-Carlo sweeps\;
}
}
Perform $N_{\rm rattle}$ Monte-Carlo sweeps\;
}
\caption{Softcore Compression Protocol\label{alg:softcore-compression}}
\end{algorithm}

The softcore protocol, originally devised for hard spheres by Xu et al. \cite{Xu2005} was implemented in \CC~as well. The main idea is to use the gradient of the potential to push overlapping spheres away from each other to produce a valid hard sphere state from a state with local violations of the constraint. The authors then use alternating compression and expansion to produce states where spheres are in close contact with each other. 

We implement a compression step closely analogous to Ref. \onlinecite{Xu2005} for our hardcore spin model as follows, given as pseudocode in \autoref{alg:compression}. Consider a valid initial state $\mathcal S = \{\vec S_i\}$ at some initial exclusion angle $\Delta$, that is $V^{(\rm tot)}(\mathcal S, \Delta) = \sum_{\expval{ij}} V(\phi_{ij}, \Delta) < \epsilon$ for some fixed tolerance $\epsilon$, which we always take to be $\epsilon = 2 \times 10^{-10}$.
Now, starting with compression, increase the inclusion angle $\Delta$ by some trial value $\delta > 0$, and use conjugate gradient descent to get from the state $\mathcal S$, which has now a finite potential energy $V^{(\rm tot)}(\mathcal S, \Delta)$, to a final state $\mathcal S'$. 
Depending on the value of $V^{(\rm tot)}(\mathcal S', \Delta)$, we continue with either compression or expansion of the system. If $V^{(\rm tot)}(\mathcal S', \Delta) > \epsilon$, we revert the state of the simulation to its state before the energy minimization and reduce the exclusion angle $\Delta$ by $\delta$ (expansion). If $V^{(\rm tot)}(\mathcal S', \Delta) < \epsilon$, we increase the exclusion angle $\Delta$ by $\delta$ (compression). In both cases, we then minimize the potential energy again by conjugate gradient descent and repeat the procedure. Every time we switch from compression to expansion or vice versa, the increment $\delta$ is halved. The step terminates if $\epsilon/2 < V^{(\rm tot)}(\mathcal S, \Delta) < \epsilon$. Note that although we use the softened constraint in principle, forcing the system to have zero potential energy is equivalent to imposing the hardcore constraint. Allowing the potential energy to be small ($\epsilon \ll N^{-1}$) but nonzero then ensures that the hardcore constraint is almost fulfilled by the state.

Similar to the LS protocol (\autoref{alg:LS-compression}), we augment the compression step with local Monte-Carlo dynamics as follows.
If the difference between successive exclusion angles $\Delta$ is below some threshold $\delta_{\rm min}$, we reduce the exclusion angle $\Delta$ by $N_{\rm exp} \cdot \delta_{\rm exp}$, where $N_{\rm exp}$ is the number of expansion already performed since the last successful update of $\Delta$, before applying $N_{\rm rattle}^{\rm (exp)}$ Monte-Carlo sweeps and applying the compression step (\autoref{alg:compression}) again. We repeat this procedure with increasing $\delta_{\rm exp}$ until either the exclusion angle is finally increased by more than the threshold or we terminate the procedure after a maximum number of expansions $N_{\rm exp}^{(\rm max)}$. The results of the LS protocol (\autoref{alg:LS-compression}) can be reproduced by setting the trial increment of the compression step $\delta_0$ to a small value $\delta_0 = 10^{-5}$, and by introducing some extra rattling sweeps $N_{\rm rattle}$ between successful increments. These extra rattling sweeps then set the timescale of compression, as $N_{\rm rattle}$ does in the LS protocol. The protocol is given as pseudocode in \autoref{alg:softcore-compression}. Here and in the main text, whenever we refer to states produced by the softcore compression protocol, we use $N_{\rm rattle}^{\rm (exp)}=10$, $N_{\rm exp} = 100$, $\delta_{\rm exp} = 5 \times 10^{-4}$, $\delta_{\rm min} = 10^{-6}$, $\delta_{\rm max} = 0.05$ and $N_{\rm rattle} = 0$.

\subsection{Statistical Properties of final states.}

\begin{figure}
    \centering
    \includegraphics{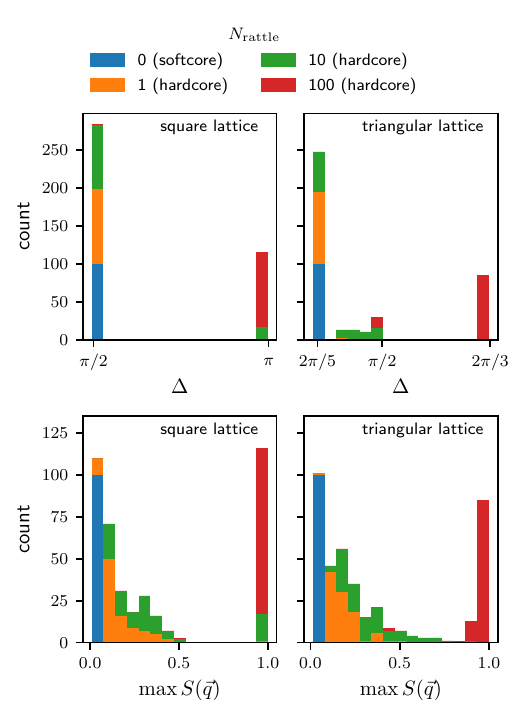}
    \caption{Properties of the final states of the compression protocols. Top: Histogram of exclusion angles at which the different protocols (indicated by color) fail on the square (left) and triangular (right) lattice. The number of rattling sweeps $N_{\rm rattle}$ sets the timescale for compression, which is faster for smaller $N_{\rm rattle}$. The values $N_{\rm rattle} = 1,\,10,\,100$ are used for the LS protocol (\autoref{alg:LS-compression}) while $N_{\rm rattle} = 0$ is used for the softcore protocol (\autoref{alg:softcore-compression}). The latter protocol utilizes a softened version of the constraint to compress the system faster and thus evade partial equilibration.  Bottom: Histogram of the maximum value of the static structure factor $S(\vec q)$. For all protocols, we performed 100 runs on a $L=40$ square and $L=42$ triangular lattice.}
    \label{fig:hists}
\end{figure}

In \autoref{fig:hists}, we show histograms of properties of the final states of the two protocols and in the case of the LS protocol also for different $N_{\rm rattle}$ (indicated by color). 
In the top row, we show the final exclusion angle, and in the bottom row, we show the maximum value of the structure factor in reciprocal space.
Since the LS protocol can only realize moderate compression speeds, the system partially equilibrates. This is reminiscent of challenge that arises for monodisperse disks, for which the protocol finds polycrystalline packings that do not pass muster as maximally random jammed states~\cite{Lubachevsky1990,Atkinson2014}. 
Note that the even though the system in the thermodynamic limit is not long range ordered, for any finite system size the maximum of the structure factor has a finite equilibrium expectation value. In particular, using the reflect algorithm developed in \cite{Sommers2020}, we find ${\rm max}\left[S(\vec q)\right] \approx 0.450$ (0.392) for the $L = 40$ ($L=80$) square lattice at $\Delta = \pi/2$ and ${\rm max}\left[S(\vec q)\right] \approx 0.384$ at $\Delta = 2\pi/5$ on the $L = 42$ triangular lattice. With sufficiently rapid compression, particularly with the softcore protocol, the order parameter stays well below this equilibrium value.

%

\section{Measuring the defect mobility}

\begin{figure}
    \centering
    \includegraphics[width=0.49\textwidth]{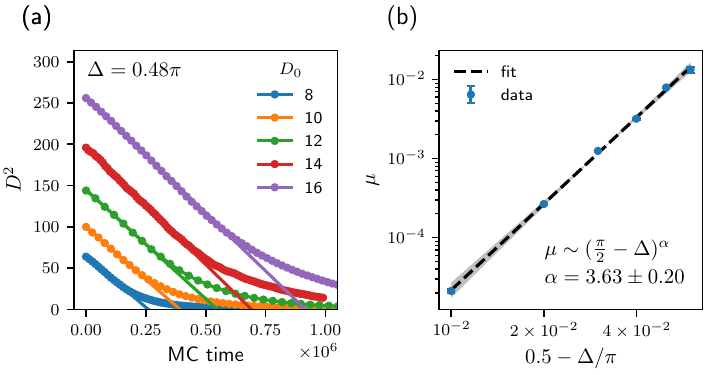}
    \caption{(a) average distance $D$ of a single vortex-antivortex pair prepared on a $L=60$ square lattice with an initial distance $D_0$. Time evolution is performed at exclusion angle $\Delta=0.48\pi$. For large distances/short times, the time dependence is well described by $D(t) = \sqrt{D_0^2 - \mu t}$. Results are averaged over $10^4$ initial states. (b) the mobility $\mu$ as a function of $\Delta$ (\autoref{fig:square} (d) of the main text).}
    \label{fig:mobility}
\end{figure}

In this appendix, we explain how the mobility of vortices, shown in~\autoref{fig:square} (d) of the main text as a function of exclusion angle $\Delta$, is measured. 

In the KT phase, a vortex-antivortex pair at distance $D$ is subject to an entropic attractive interaction potential $V(D) \sim \log(D)$, which makes the pair eventually annihilate. For a single pair, initially at distance $D_0$, the distance as a function of time should then satisfy the differential equation
\begin{equation}
    \dot D(t) = \mu\,F[D(t)] =-\frac{\mu}{D(t)}
\end{equation}
where $F(D)$ is the attractive force between the pair and $\mu$ is the mobility of vortices. The above equation has a simple solution
\begin{equation}
    D(t) = \sqrt{D_0 - \mu t}.
    \label{eq:distance-time}
\end{equation}
We compute the mobility as a function of exclusion angle $\Delta$ numerically as follows. For each value of $\Delta$, we prepare single vortex-antivortex pairs at distance $D_0$ on a $L=60$ lattice, by placing it into a N\'eel state and evolving the state using $10^6$ Monte-Carlo sweeps, but fixing the vortex and antivortex in place.
After this, we evolve the system further under the Monte-Carlo dynamics, but without fixing the vortex-antivortex pair. While doing so, we keep track of the distance of the pair as a function of time.
We show the result of this procedure in \autoref{fig:mobility}. In (a), we show the distance between the vortex antivortex pair as a function of Monte-Carlo time, averaged over $10^4$ initial states, for a range of initial distances at $\Delta = 0.48\pi$. As expected, for short times and long distances, it obeys the functional form of \autoref{eq:distance-time}. 
The mobility can then be extracted from a linear fit to $D^2(t)$ and is shown as a function of the exclusion angle in (b). 

We have also measured the defect mobility for ``Hamiltonian'' dynamics, results of which are presented in \autoref{app:hamiltonian-vertex-mob}.

%

\section{Equilibrium properties}

In this appendix, we provide additional details on the computation of equilibrium properties of hardcore spins which go beyond those already presented in Ref. \onlinecite{Sommers2020}.

\subsection{The vortex density on the square lattice vanishes as a power law within the KT phase\label{app:eq_vorticity}}

\begin{figure}
    \centering
    \includegraphics{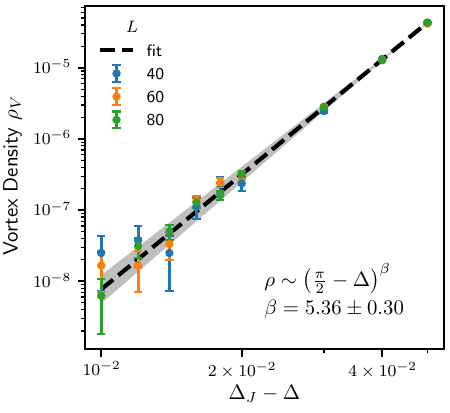}
    \caption{Equilibrium vortex density $\rho$ inside the Kosterlitz-Thouless phase. The results are compatible with the fact that $\rho$ vanishes as a power law as $\Delta \to \Delta_J$. The area between the two lines with exponent $\beta \pm \sigma_\beta$ is indicated in gray.}
    \label{fig:eq_vorticity}
\end{figure}

In the main text, we described how the mechanism of jamming of hardcore spins is intimately related to topological defects. In particular, these defects become strictly forbidden at an exclusion angle $\Delta_J$ that is the defect density $\rho$ must vanish as $\Delta \to \Delta_J$ from below. In this section, we investigate how this happens.

Focusing on the square lattice model, vortices become forbidden at $\Delta_J > \Delta_{\rm KT}$, that is inside the Kosterlitz-Thouless (KT) phase. In the range $\Delta_{\rm KT} < \Delta < \Delta_J$, vortices are already confined, but fluctuations lead to a finite density
\begin{equation}
    \rho_V = \frac{1}{N_p}\sum_p \abs{V_p}
\end{equation}
where the sum is over all plaquettes $p$ in the lattice, $N_p$ is the number of plaquettes ($N_p = N$ on the square lattice with periodic boundaries), and $V_p$ is the vorticity of the plaquette $p$
\begin{align}
    V_{p = (i,j,k,l)} = \frac{1}{2\pi}&\left[\measuredangle \left(\tilde{\vec S}_i, \tilde{\vec S}_j \right) +
    \measuredangle \left( \tilde{\vec S}_j, \tilde{\vec S}_k \right)\right. \notag \\ 
    &+ 
    \left.\measuredangle \left( \tilde{\vec S}_k, \tilde{\vec S}_j \right) +
    \measuredangle \left( \tilde{\vec S}_l, \tilde{\vec S}_i \right)\right]
    \label{eq:vorticity-sq}
\end{align}
where $\measuredangle \left( \tilde{\vec S}_i, \tilde{\vec S}_j \right) \in [-\pi, \pi]$ is the signed phase difference between the spins $\tilde{\vec S}_i$ and $\tilde{\vec S}_j$ and the tilde denotes the fact that the spins are taken with respect to the N\'eel state.
The vortex density in equilibrium on the square lattice is shown as a function of exlcusion angle $\Delta$ in \autoref{fig:eq_vorticity}. The results are compatible with a power law, that is 
\begin{equation}
    \rho_V \sim \left(\Delta_J - \Delta \right)^\beta
\end{equation}
with $\Delta_J = \pi/2$ and $\beta \approx 5.36 \pm 0.30$.

The fact that the defect density $\rho$ vanishes continuously (rather than via a first order transition) has nontrivial consequences also for the compression dynamics. As explained in the main text, slow compression is able to reach the exclusion angle $\Delta_{\rm max}$. This is because under slow compression, the system then stays in equilibrium and, because the defect density vanishes continuously as $\Delta \to \Delta_J$ from below, has no defects when reaching $\Delta_J$, where defects become strictly forbidden by the hardcore constraint. The protocol can then compress beyond that point, in contrast to a fast compression protocol which falls out of equilibrium and reaches $\Delta_J$ with a nonzero density of defects.

\subsection{Forbidden defects on the triangular lattice}

\begin{figure}
    \centering
    \includegraphics{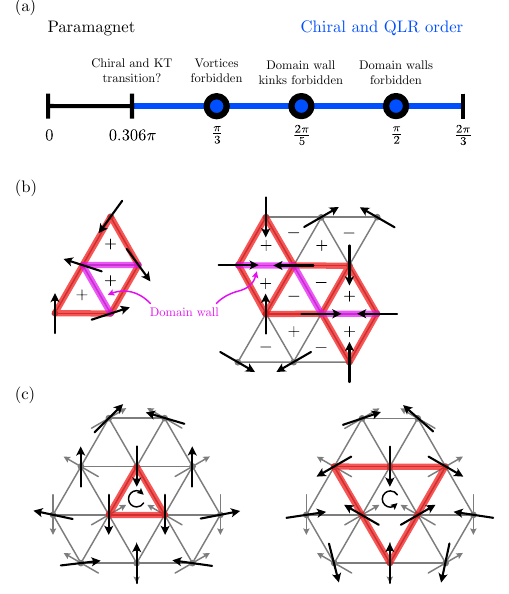}
    \caption{Forbidden defects and phase diagram on the triangular lattice. (a) Vortex (right) and Antivortex (left) with respect to one of the two possible 120-degree states. The 120 degree state is indicated in gray. The vortices and antivortices become strictly forbidden at $\Delta > \pi/3$. Here and in (b), we indicate the edges that minimize the enclosed angle in red. 
    (b) Domain walls of chirality. The sign of the vector chirality is indicated as $\pm$ in the triangles. In the two 120 degree states, up and down pointing triangles have opposite chirality, we indicate domain walls between the two possibilities in purple. Smooth domain walls (right) become strictly forbidden at $\Delta > \pi/2$, while ``kinks'' in domain walls (left) become forbidden already at $\Delta > 2\pi/5$.
    (c) Equilibrium phase diagram. Within the precision of our numerical estimate, the transitions to chiral and quasi long range (QLR) order coincide. We also indicate the succession of forbidden defects. The system exhibits extremely slow relaxation dynamics from $2\pi/5$, with the dynamical transition thought to occur as $\Delta \to \pi/2$.}
    \label{fig:equil-phases-tri}
\end{figure}

The phase diagram of hardcore spins on the triangular lattice is shown in \autoref{fig:equil-phases-tri} (a). As on the square lattice, there is a low-exclusion-angle paramagnetic and a large-exclusion-angle quasi-long-range ordered phase. However, the situation on the triangular lattice is richer because in addition to the $O(2)$ symmetry of the XY spins, there is a chiral ($\mathbb{Z}_2$) symmetry which the noncollinear ground state of the model breaks. Because of this, there are two kind of topological defects: vortices as well as domain walls.
The former are defined with respect to a state of fixed chirality, whose noncollinearity leads to vortices and antivortices having very different free energy costs associated with their creation.
The latter even become incompatible with the hardcore constraint at different exclusion angles depending on the their local shape [\autoref{fig:equil-phases-tri} (b)]. 

To understand all of this in detail, we must get a sense of what kind of spin textures the topological defects correspond to. 
We begin by considering domain walls of chirality. For this, note that at $\Delta_{\rm max} = 2\pi/3$, not all allowed states can be transformed into each other via global rotations. Instead, there are two incarnations of the  ``120-degree'' or ``$\sqrt 3 \times \sqrt 3$'' state, which are related by the exchange of two sublattices. These two states are shown in the top right and bottom left of the right of \autoref{fig:equil-phases-tri} (b). There, we also show how they can be separated by a domain wall, indicated in pink. To distinguish the two domains quantitatively, we define a vector chirality $\kappa_t$ on each elementary triangle $t$ of the lattice
\begin{equation}
    \kappa_{t = (i,j,k)} = \vec S_i \cross \vec S_j + \vec S_j \cross \vec S_k + \vec S_k \cross \vec S_i.
    \label{eq:chirality}
\end{equation}
In the 120 degree state, the chirality is extremal on all triangles, that is $\abs{\kappa_t} = 3 \sqrt{3}/2$, and it has opposite sign on upwards and downwards triangles. This is also shown in \autoref{fig:equil-phases-tri} (b), where we indicate the sign of $\kappa$ for each triangle as $+$ or $-$. We call a domain wall as shown on the right of \autoref{fig:equil-phases-tri} (b) "smooth," because the domain wall takes no sharp turns or ``kinks'' as shown on the left of the same figure. Inspecting the smooth configuration, we recognize the 90-degree structure familiar from the vortex cores on the square lattice. We thus conclude that these kind of domain walls become forbidden at $\Delta = \pi/2$. 
Turning to the domain wall kink, note that such a configuration necessitates three adjacent triangles to have the same sign of the vector chirality $\kappa$. This however already implies that the winding number along the loop encircling the three triangles is $2\pi$, which becomes incompatible with the hardcore constraint for $\Delta > 2\pi/5$.

Turning to vortices, these are defined exactly as on the square lattice, but with respect to a 120-degree state of fixed chirality. In particular we define a vorticity $V_t$ 
\begin{align}
    V_{t = (i,j,k)} = \frac{1}{2\pi}\left[ \measuredangle \left(\tilde{\vec S}_i, \tilde{\vec S}_j \right) +
    \measuredangle \left( \tilde{\vec S}_j, \tilde{\vec S}_k \right) + 
    \measuredangle \left( \tilde{\vec S}_k, \tilde{\vec S}_i \right) \right]
    \label{eq:vorticity}
\end{align}
where $\measuredangle \left( \tilde{\vec S}_i, \tilde{\vec S}_j \right) \in [-\pi, \pi]$ is the signed phase difference between the spins $\tilde{\vec S}_i$ and $\tilde{\vec S}_j$ and the tilde denotes the fact that the spins are taken with respect to one of the 120-degree states. Note that this implies that vortices are only well defined in the presence of chiral order. For a uniform choice of chirality, we show a vortex (right) and antivortex (left) in \autoref{fig:equil-phases-tri} (c). As in (b), we highlight the bonds that minimize the enclosed angle between spins in red. 
In contrast to the square lattice case, the spin textures of vortices and antivortices on the triangular lattice are qualitatively different. Still, as we will show in the following, both become forbidden by the hardcore constraint at the same exclusion angle, that is $\Delta = \pi/3$. 

First, note that a perfect (meaning unperturbed) antivortex core is incompatible with the hardcore constraint for any nonzero exclusion angle. What is still possible is a state in which the configuration on the center triangle of the antivortex is perturbed. For this, two of its spins are rotated in opposite directions, away from the third. For the center triangle on the right of \autoref{fig:equil-phases-tri} (c), one possibility of this would be to rotate the bottom right spin of the triangle clockwise and the bottom left spin anticlockwise. 
Via such a rotation by an angle $\epsilon$, we can produce a valid configuration on the center triangle for any $\epsilon < \Delta$. However the vorticity on the central triangle for such state is only equal to $-1$ as long as $\abs{\epsilon} < \pi/3$, which means that for $\abs{\epsilon} > \pi/3$ it is not an antivortex core. Since an antivortex needs a core, its existence is hence only compatible with the hardcore constraint for $\Delta < \pi/3$.

Second, the vortex core as constructed on the right side of~\autoref{fig:equil-phases-tri} (c) is just the 120-degree configuration of opposite chirality, which might seem to suggest compatibility with the hardcore constraint for all exclusion angles up to $\Delta_{\rm max}$. However, a vortex is not just a core but an extended object. The next largest closed loop to consider is of length six and indicated in red on the left of \autoref{fig:equil-phases-tri} (c). The spin configuration of this loop in a perfect vortex is clearly only compatible with the hardcore constraint for $\Delta < 2\pi/6 = \pi/3$. At this point, only three spins with any available phase space are those at the three corners, which however can only be deformed in a way that already flips the vorticity of the whole loop.

The exclusion angles at which the different topological defects become forbidden by the hardcore constraint are summarized in \autoref{fig:equil-phases-tri} (a).

\subsection{Details on the numerical estimation of $\Delta_{\rm KT}$ on the triangular lattice\label{app:eq_tri}}

Here, we give the details on our numerical computation of the equilibrium phase diagram of hardcore spins on the triangular lattice. We use a variant of the Wolff algorithm, in which a single Monte-Carlo step consists of a global reflection of a cluster of spins. The axis of reflection is chosen at random and a "pocket" is initialized to contain a random seed spin. Each spin in the pocket is reflected about the chosen axis, and if doing so leads to a violation of the hard constraint with its neighbors, these neighbors are added to the pocket. The move terminates when the pocket is empty, i.e., when the lattice has returned to an allowed spin configuration. By definition, the ``cluster'' consists of all spins that are added to the pocket at some point in the move. We call this the \texttt{reflect} algorithm.

While the \texttt{reflect} algorithm is provably ergodic on bipartite lattices \cite{Sommers2020}, we consider it likely that the triangular lattice hosts exceptions. One indication of this is that the average cluster size near the critical point scales as $L^2$, whereas the square lattice enjoys a superior scaling of $L^{2-\eta'}$, with exponent $\eta'$ in close agreement with the critical exponent $\eta$ of the staggered spin susceptibility~\cite{Sommers2020}. Nevertheless, several other metrics---including the decay of the autocorrelation function of the spin susceptibility, the ability of cluster moves to unwind vortices and extended defects, and the convergence of different initial states to the same time-averaged expectation values---support the use of the algorithm as an approximate probe of the equilibrium phase diagram, especially at low $\Delta$.

Based on studies of the thermal model \cite{Obuchi2012}, the spin and chiral transitions are expected to at very close exclusion angles, at or below $\pi/3$, where vortices become forbidden. Finite-size scaling on lattices up to $L=54$ at angles ranging from $0.3\pi$ to $0.4\pi$ support this hypothesis, suggesting a phase transition around $0.307\pi$.

\begin{figure}
    \centering
    \includegraphics[width=\linewidth]{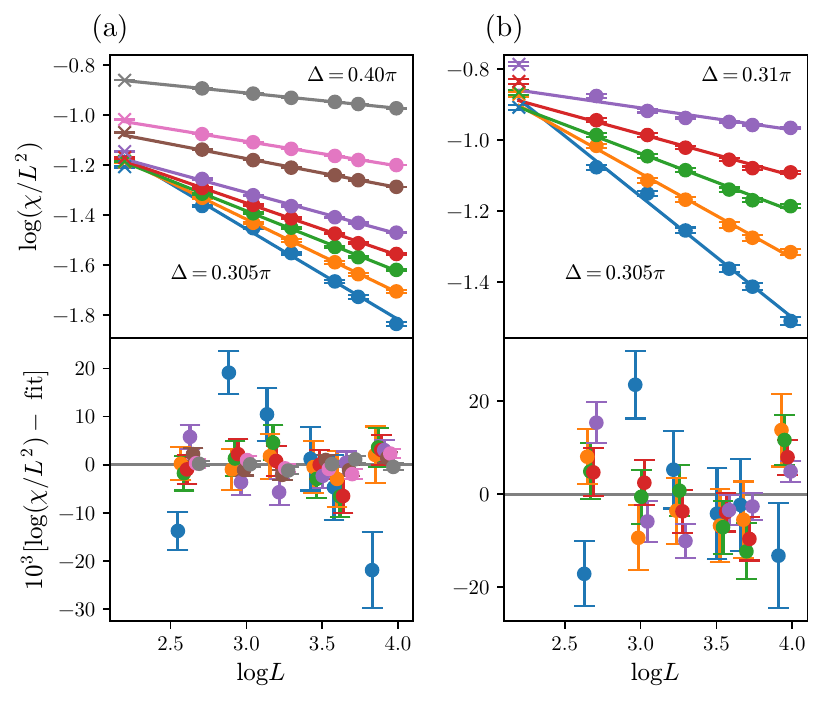}
    \caption{Finite-size scaling of spin and chiral susceptibilities for system sizes ranging from $L=9$ to $L=54$. Only system sizes with $L\geq 15$ are included in the power-law fits, points excluded from the fit are indicated by crosses rather than circles. Residuals with respect to the fit, amplified by a factor of $10^3$, are shown in the bottom subplots. (a) Spin susceptibility,  defined in ~\autoref{eq:susceptibility}, for exclusion angles $\Delta=0.305\pi,0.306\pi,0.307\pi,0.308\pi,0.31\pi,0.32\pi,0.33\pi,0.4\pi$. The strong concave-down trend in the residuals for $\Delta=0.305\pi$ indicates a poor power-law fit, as this exclusion angle is in the paramagnetic phase. (b) Chiral susceptibility, defined analogously with~\autoref{eq:chirality} in place of the magnetization, for exclusion angles ranging from $0.305\pi$ to $0.31\pi$, in the vicinity of the continuous phase transition. In addition to a concave-down trend in the residuals for $\Delta=0.305\pi$, which lies in the chirally disordered phase, there is a clear concave-up trend in the residuals on the other side of the transition. \label{fig:scale}}
\end{figure}

Looking first at the spin transition, since the associated $U(1)$ symmetry is continuous, only QLRO is permitted for $\Delta < \Delta_{max}$, with a continuously varying exponent $\eta$ in the critical phase. This exponent is determined from the scaling of the spin susceptibility, or structure factor, defined as the maximum in the Fourier spectrum of the spin correlation function:
\begin{equation}\label{eq:susceptibility}
    S(\vec{q}) = \sum_{\vec{r}} e^{i\vec{q}\cdot \vec{r}} \langle \vec{S}(\vec{0})\vec{S}(\vec{r})\rangle 
\end{equation}
Triangular lattice exclusion models in the QLRO phase exhibit a peak at $\vec{q_s}=(4\pi/3,0)$ corresponding to the ``120-degree'' or ``$\sqrt{3}\times\sqrt{3}$'' order. The susceptibility $\chi_s = S(4\pi/3,0)$ then scales as a power law with the finite system size $\chi_s \sim L^{2-\eta}$, as shown in~\autoref{fig:scale} (a).

\begin{figure}
    \centering
    \includegraphics[width=\linewidth]{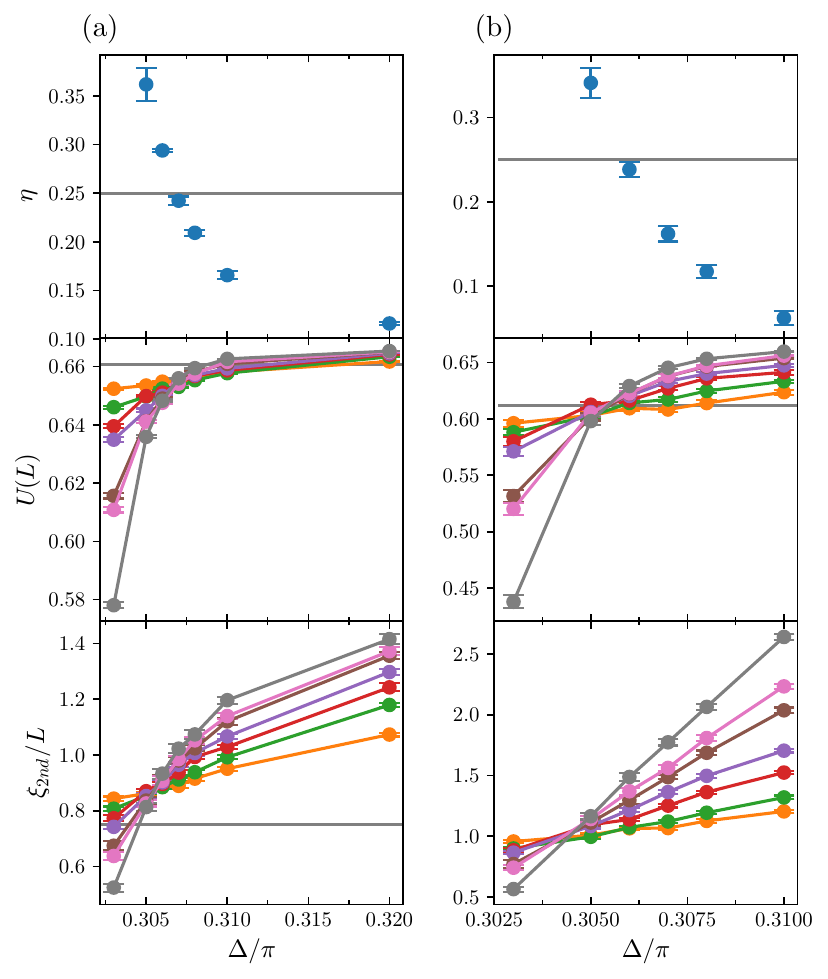}
    \caption{Anomalous critical exponent, Binder cumulant, and correlation length for near the transition. (a)  Spin observables, with horizontal lines indicating universal values for the KT universality class~\cite{Hasenbusch2005,Hasenbusch2008}. (b) Chiral observables, with horizontal lines indicating values for the Ising universality class on the triangular lattice~\cite{Kamieniarz1993}.}
    
    \label{fig:critical-KT}
    
\end{figure}

If the spin transition belongs to the Kosterlitz-Thouless universality class, the critical exponent will assume the value $\eta=1/4$ at the critical angle $\Delta_{\rm KT}$. Precision numerical studies are split on whether the magnetic transition in this model and the closely related fully frustrated XY model is in the KT universality class~\cite{Lee1998,Hasenbusch2005b,Lv2012} or of a non-KT character~\cite{Okumura2011, Obuchi2012}. At the system sizes considered here we cannot draw a firm conclusion regarding the universality class for the constraint-only model, but the data in~\autoref{fig:critical-KT} (a) are broadly consistent with a KT transition. First, $\eta_s=1/4$ (the value at the KT transition) around $\Delta\approx 0.307\pi$, which is also the angle below which the power-law scaling of $\chi_s$ becomes a poor fit, indicating a transition to the paramagnetic phase. In this neighborhood of $\Delta$, we also examine the Binder ratio (middle panel), which should also take a universal value at the critical point, and become asymptotically independent of $L$ above this angle. A variety of normalizations exist in the literature, but we use the definition~\cite{Hasenbusch2008}:
\begin{equation}\label{eq:binder}
U(L) = 1 - \langle |\vec{M_s}|^4 \rangle/3\langle |\vec{M_s}|^2\rangle ^2
\end{equation}
where $\vec{M}_s$ is the K-point magnetization~\cite{Obuchi2012}. Finally, we measure the second-moment correlation length (bottom panel), defined as~\cite{Obuchi2012}:
\begin{equation}\label{eq:xi2}
\xi_{2nd}(L) = \frac{1}{2\sin(\pi/L)}\sqrt{S(\vec{q_s})/S(\vec{q_s}+\frac{2\pi}{L}\vec{\hat{x}}) - 1}.
\end{equation}
To leading order, $\xi_{2nd}/L$ becomes independent of system size in the QLRO phase and takes a universal value in the thermodynamic limit at the KT transition~\cite{Hasenbusch2005}.

Turning to the chiral phase transition, we measure analogous quantities to Eqs.~(\ref{eq:susceptibility}),~(\ref{eq:binder}) and~(\ref{eq:xi2}), with the chirality (\autoref{eq:chirality}) serving as the order parameter in place of the K-point magnetization. In this case, since the associated symmetry is discrete, the transition is to a long-range-ordered phase. Consequently, in the thermodynamic limit, the connected correlation function is algebraically decaying only precisely at the critical point. In terms of the finite-size scaling of the chiral susceptibility, this means that the power-law scaling with $L$ is only exhibited in the close vicinity of the transition, as shown in~\autoref{fig:scale} (b). The system-size independence of the Binder ratio and second-moment correlation length, as an additional signature of critical behavior, also appears only in the vicinity of the transition. These signatures are shown in~\autoref{fig:critical-KT} (b), again indicating a critical point in the range of $\Delta\approx (0.306\pm0.001)\pi$. Consistent with the expectation on grounds of symmetry that the chiral transition belongs to the Ising universality class, the value of $\eta$ deduced from a power-law fit to $\chi$ is close to $\eta_{Ising}=1/4$. Moreover, the value of $U_c$ attained at the intersection of different system sizes is consistent with the Binder ratio at the Ising critical point on a triangular lattice geometry~\cite{Kamieniarz1993}, a universal value that has also been observed at the chiral transition of the antiferromagnetic XY model~\cite{Lv2012}.

To summarize, within uncertainties, the data are suggestive of coincident chiral and spin transitions, with the latter falling in the KT universality class. The possibility of two separate transitions cannot be ruled out; in the thermal model, the difference between the two transitions only manifests above a threshold system size of $L\approx 300$, so the same effect may be present here~\cite{Obuchi2012}. Regardless, the existence of a transition below $\Delta=\pi/3$ indicates that, as on the square lattice, there exists a range of exclusion angles where defects are geometrically allowed but entropically disfavored, leading to quasi-long-ranged order by disorder. In terms of the dynamics, this means that the jamming point and its concomitant doubly algebraic relaxation dynamics occur within the QLRO phase.

%

\section{Hamiltonian Dynamics\label{app:hamiltonian-dynamics}}

Here we outline the main results of the study of Hamiltonian dynamics on the square lattice. In these dynamics (\autoref{alg:hamiltonian}), each spin is initialized with an angular velocity drawn from a Gaussian distribution, and we work in the frame in which the net angular momentum is zero. The system is evolved until a pair of neighboring spins "collides," i.e. saturates the hard constraint, at which point the two colliding spins swap velocities.

\begin{algorithm}
\SetAlgoLined
\SetKwInOut{Output}{Output}
\SetKwInOut{Input}{Input}
\Input{Valid configuration $\mathcal S_0 = \{\vec{S}_i\}$ at $\Delta+\epsilon$\\
Velocities $\{\omega_i\}$ sampled from Gaussian distribution}
\Output{Valid configuration $\{\vec{S}_i\}$ at $\Delta$, evolved to time $T$}
$\mathcal S \gets \mathcal S_0$\;
$t \gets 0$;
$\mathcal{P}\gets$ empty priority queue of future collisions\;
\For{($\vec{S}_j, \vec{S}_k$) \normalfont{\textbf{in} nearest neighbor pairs of} $\mathcal{L}$} {
$t_{coll}\gets$ future collision time between $\vec{S}_j$, $\vec{S}_k$\;
add $(t_{coll}, \vec{S}_j, \vec{S}_k, \omega_j, \omega_k)$ to $\mathcal{P}$\;
}
\While{$t<T$}{
$(t_{coll}, \vec{S}_j, \vec{S}_k, \omega_j, \omega_k) \gets$ pop $\mathcal{P}$\;
$t_f \gets \min(t_{coll}, T)$\;
\For{$\vec{S}_i$ in $\mathcal{S}$} {
$\phi_i \gets \phi_i + (t_f-t)\omega_i$\;
}
swap $\omega_j$ and $\omega_k$\;
Update priority queue with new collision times\;
$t \gets t_f$\;
}
\Return $\mathcal S$;
\caption{Hamiltonian Dynamics\label{alg:hamiltonian}}
\end{algorithm}
\subsection{Conservation Laws}
The Hamiltonian dynamics are more constrained than the stochastic Monte Carlo dynamics in a few crucial ways. 

First, the vorticity on each plaquette with respect to the antiferromagnetic state,
\begin{align}
    V_{p = (i,j,k,l)} = \frac{1}{2\pi}&\left[\measuredangle \left(\tilde{\vec S}_i, \tilde{\vec S}_j \right) +
    \measuredangle \left( \tilde{\vec S}_j, \tilde{\vec S}_k \right)\right. \notag \\ 
    &+ 
    \left.\measuredangle \left( \tilde{\vec S}_k, \tilde{\vec S}_j \right) +
    \measuredangle \left( \tilde{\vec S}_l, \tilde{\vec S}_i \right)\right]
    \label{eq:vorticity}
\end{align}
is conserved. To see that this is true, denote the four spins belonging to a plaquette as $\phi_0, \phi_1, \phi_2,\phi_3 \in [0,2\pi)$, where without loss of generality we can define the $x$ axis such that $\phi_0=0$ at a given instant in time. 
\comment{
\begin{subequations}\label{eq:in-ex}
\begin{align}
\phi_0 &= \tilde{\phi}_0 = 0 \\
\phi_1 &= (\tilde{\phi}_1 + \pi) \mod 2\pi \\
\phi_2 &= \tilde{\phi}_2 \\
\phi_3 &= (\tilde{\phi_3} + \pi) \mod 2\pi
\end{align}
\end{subequations}
}
Denoting spin orientations with respect to the Ne\'el state by $\tilde{\vec{S}}_j=(-1)^{j}\vec{S}_j$, the enclosed angle between $\tilde{\vec{S}}$ and $\tilde{\vec{S}}'$ is defined in the interval $[-\pi, \pi]$:
\begin{equation}\label{eq:ang}
\measuredangle(\tilde{\vec{S}}, \tilde{\vec{S}}') = (\tilde{\phi} - \tilde{\phi}' + \pi) \mod  2\pi - \pi.
\end{equation}
Plugging this into the expression for the vorticity yields:
\begin{align}
V_{p=0123} &= \frac{1}{2\pi}\big[(\tilde{\phi}_1 + \pi) \mod 2\pi - (\tilde{\phi}_1 + \pi - \tilde{\phi}_2) \mod 2\pi \notag \\
&+ (\tilde{\phi}_3 + \pi - \tilde{\phi}_2) \mod 2\pi - (\tilde{\phi}_3 + \pi) \mod 2\pi\big] \notag \\
&= -H(\phi_2 - \phi_1) + H(\phi_2 - \phi_3)
\end{align}
where $H(x)$ is the Heaviside function. Since the relative ordering of $\phi_1, \phi_2$ and $\phi_2, \phi_3$ is preserved by the dynamics, $V_p(t)$ depends only on the initial conditions.
\comment{
\begin{enumerate}
\item If $\theta_1 < \theta_2 < \theta_3$, then $\Phi=-2\pi$, an antivortex.
\item If $\theta_1 > \theta_2 > \theta_3$,  then $\Phi =  2\pi$, a vortex.
\item If $\theta_1, \theta_3 < \theta_2$, then $\Phi = -2\pi + 2\pi = 0$. Similarly if $\theta_1, \theta_3 > \theta_2$, then $\Phi = 0$.
\end{enumerate}
}

Thus,  there is one local conserved quantity per plaquette. A square lattice of $N$ spins has $N$ plaquettes, so this is enough to make the system integrable~\cite{Hand1998}. In addition, periodic boundary conditions impose that the net topological charge on the lattice is zero.

An additional global constraint comes from the fact that the colliding spins behave like equal-mass objects in 1D, which simply exchange velocities upon collision. This implies that the set of velocities is fixed by initial conditions. In the problem of hard rods confined to a 1D box, this conservation law enables an explicit solution~\cite{Jepsen1965,Percus1969}. First, positions are redefined such that the rods have zero length. Then, the positions of these point particles at later times are given by free trajectories labeled $\{x_i(t)\}$, but with each intersection of trajectories inducing a swap of the labels. However, the difference between translational degrees of freedom in real space and orientational degrees of freedom in spin space complicates attempts to apply a similar solution strategy to the problem of hardcore spins. Rather than pursuing this strategy here, in keeping with the main focus of this letter, we will lift some of the constraints to recover a form of "jamming."
\subsection{Tunneling}
Among the several quantities that exhibit power-law scaling near jamming in the Monte Carlo dynamics is the vortex mobility, which vanishes with power $\alpha$ as $\Delta\rightarrow \Delta_J$ [\autoref{fig:square}(d)]. Since~\autoref{alg:hamiltonian} imposes zero mobility at all angles, this constraint must be lifted to observe a similar scaling in the Hamiltonian dynamics. A minimal modification that achieves this end is a change to how the spins are updated in a collision. In place of line 12, in which the colliding spins exchange velocities, the update in~\autoref{alg:hamiltonian-tun} is applied: one of the two spins is chosen at random to "tunnel" through the other, i.e., is reflected about the axis of the other spin (lines 1-2). If the proposed reflection causes the tunneled spin to violate the exclusion constraint with one of its other neighbors, the move is rejected and velocities are exchanged instead (lines 5-9), as in the original algorithm.

\begin{algorithm}[hbtp]
\SetAlgoLined
\SetKwInOut{Output}{Output}
\SetKwInOut{Input}{Input}
\Input{Valid configuration $\mathcal S = \{\vec{S}_i\}$, velocities $\{\omega_i\}$\\
Neighboring spins $(\vec{S}_j, \vec{S}_k) : |\measuredangle(\vec{S}_j, \vec{S}_k)|=\Delta$}
\Output{Valid configuration $\{\vec{S}_i\}$, post-collision}
$(a, b) \gets$ random permutation of $(j, k)$\;
$\vec{S}_a^{old} \gets \vec{S}_a$ \;
$\vec{S}_a \gets 2(\vec{S}_a \cdot \vec{S}_b)\vec{S}_b- \vec{S}_a$\; 
\For{$\vec{S}$ \normalfont{\textbf{in} neighbors of} $\vec{S}_a$}{\If{$|\measuredangle(\vec{S}_a,\vec{S})| < \Delta + \epsilon$}{
$\vec{S}_a \gets \vec{S}_a^{old}$ \;
Swap $\omega_j$ and $\omega_k$\;
break\;}
}
\Return $\mathcal S$, $\{\omega_i\}$;
\caption{Hamiltonian Dynamics Tunneling Step\label{alg:hamiltonian-tun}}
\end{algorithm}

The reflection flips the sign of the phase difference between $\vec{S}_j$ and $\vec{S}_k$, so by definition this tunneling move creates a vortex-antivortex pair on the two plaquettes that have $(i,j)$ as common vertices. For exclusion angles above $\Delta>\pi/2$, defects are forbidden, so the tunneling moves are rejected with probability one. Likewise, in stochastic Monte Carlo dynamics, if a proposed spin flip $\vec{S}_i \rightarrow \vec{S}_i'$ induces a change in the sign of the phase difference with a neighboring spin, then it is guaranteed to violate the hard constraint for $\Delta>\pi/2$. In both cases, the dynamics just move through the sector with zero vorticity on each plaquette.

For $\Delta<\pi/2$, on the other hand, tunneling moves are accepted with nonzero probability. As the exclusion angle approaches the $\Delta_J=\pi/2$ from below, the acceptance probability vanishes, which we expect to be accompanied by a vanishing of the vortex mobility.

\subsection{Vortex mobility and diverging timescale\label{app:hamiltonian-vertex-mob}}

\begin{figure}
    \centering
    \includegraphics[width=\linewidth]{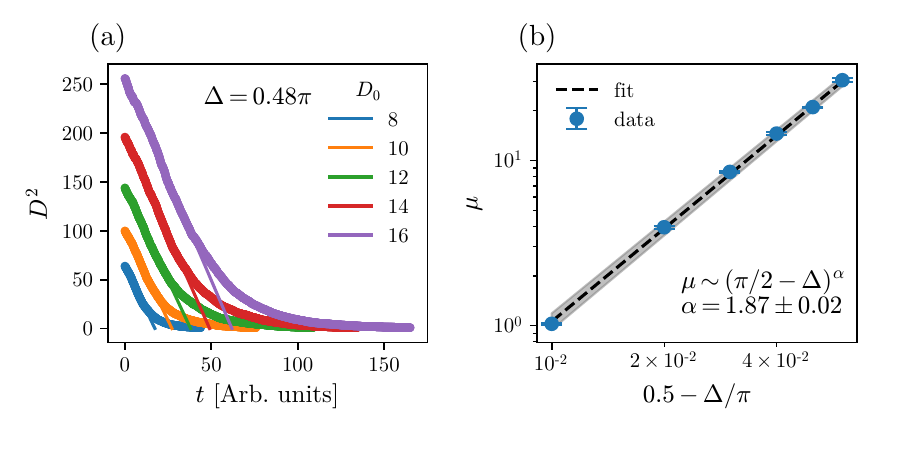}
    \caption{(a) average distance $D$ of a single vortex-antivortex pair prepared on a $L=60$ square lattice with an initial distance $D_0$. Time evolution is performed at exclusion angle $\Delta=0.48\pi$. For large distances/short times, the time dependence is well described by $D(t) = \sqrt{D_0^2 - \mu t}$. Results are averaged over $10^3$ initial states, using the same states as in the stochastic Monte Carlo dynamics. (b) the mobility $\mu$ depends on $\Delta$ and vanishes with an exponent $\alpha=1.87 \pm 0.02$ (compare~\autoref{fig:square} (d) in the main text). The area between the two lines with exponent $\alpha \pm \sigma$ is indicated in gray.}
    \label{fig:ht-mobilities}
\end{figure}

To measure the vortex mobility, as with the Monte Carlo dynamics discussed in the main text, we can measure the diffusion-annihilation process by tracking the separation of a vortex-antivortex pair as a function of time. The initial states used are the same as those in the main text, prepared by placing a single defect pair into a Ne\'el state at a distance $D_0$ and evolving the state with $10^6$ Monte Carlo sweeps. These states, whose structure factor is suppressed below the equilibrium value due to the presence of the defects, are then fed into the Hamiltonian dynamics $+$ tunneling algorithm, where the pair is no longer fixed. The results are shown in~\autoref{fig:ht-mobilities}. As anticipated from the theoretical considerations of phase-ordering kinetics, the vortex separation decays as $D^2 \propto D_0^2 - \mu t$, where the mobility $\mu$ depends only on the exclusion angle $\Delta$. As $\Delta$ approaches $\pi/2$, $\mu$ scales as a power law, but with a different exponent from the Monte-Carlo dynamics.

\begin{figure}
    \centering
    \includegraphics[width=\linewidth]{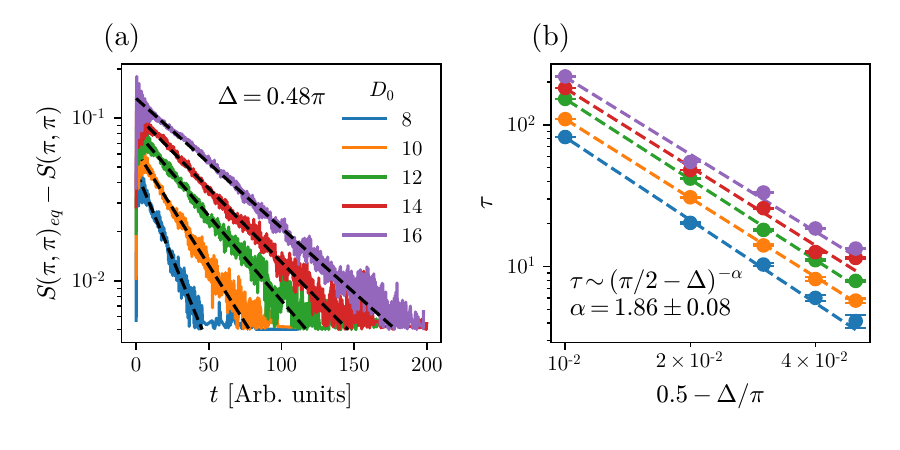}
    \caption{(a) Exponential approach of the structure factor $S(\pi,\pi)$ to its equilibrium value, for $L=60$ using the same initial states as in ~\autoref{fig:ht-mobilities}. Hamiltonian evolution with tunneling is performed at exclusion angle $\Delta=0.48\pi$, fit to an exponential $C\exp(-t/\tau)$ (black dashed lines). (b) The relaxation time $\tau$ behaves as a power law in $\pi/2-\Delta$, with the same exponent as the mobility. Consistent results are obtained for different initial separations; the reported value of $\alpha$ is an average over the best fit values for each $D_0$.}
    \label{fig:ht-sf}
\end{figure}

The vanishing mobility of defects coincides with a diverging timescale for relaxation towards equilibrium. In the main text, this is probed by tracking the defect density as a function of time starting from far-from-equilibrium initial states [\autoref{fig:square}(b-c)]. For the Hamiltonian dynamics, we found the structure factor to be a more reliable probe of the diverging timescale. The structure factor approaches its equilibrium value exponentially, with a dynamical exponent $z=2$, just as for Monte Carlo dynamics. For a given system size, the relaxation time diverges as a power law $\sim (\Delta_J-\Delta)^{-\alpha}$ with a similar exponent $\alpha$ to that of the vortex mobility.  More robust results for $\alpha$ are obtained by returning to the initial states containing a single vortex-antivortex pair. After an initial transient, $S(\pi,\pi)_{eq} - S(\pi,\pi)(t)$ exponentially decays with a relaxation time that increases with the initial separation [\autoref{fig:ht-sf}(a)]. As a function of $\Delta_J-\Delta$ [\autoref{fig:ht-sf}(b)], $\tau$ diverges with the same exponent, within uncertainties, as the vanishing mobility; the latter exponent is estimated to be $\alpha=1.87\pm 0.02$. 

\section{Softcore compression with forbidden domain wall kinks\label{app:forbidden-kinks}}

In this appendix, we give details on and present the results of softcore compression on the triangular lattice in the absence of domain wall kinks. A domain wall kink is a configuration as on the left of \autoref{fig:equil-phases-tri} (b), that is two domain walls meet at an angle of $\pi/3$ or, equivalently, three adjacent triangles have the same vector chirality. 
We begin by describing the adjusted preparation of initial states and  subsequently discuss the results of the softcore compression protocol (\autoref{alg:softcore-compression}) on the kink free model. 
In contrast to softcore compression in the presence of kinks, we here find (\autoref{fig:kink_free_delta_J}) a broad distribution between $2\pi/5$ and $\pi/2$ of the exclusion angle $\Delta_J$ at which the protocol terminates for different runs, superimposed with pronounced peaks at rational fractions of $2\pi$. The weight of the distribution shifts towards larger $\Delta$ as $N_{\rm rattle}^{(\rm exp)}$ is increased, without ever exceeding $\Delta = \pi/2$. We interpret this as evidence for the fact that we expect the first infinite relaxation timescale of the system to occur at $\Delta = \pi/2$, where domain walls become strictly incompatible with the hardcore constraint. We attribute the presence of a broad distribution of $\Delta_J$ for small $N_{\rm rattle}^{(\rm exp)}$ to the physics of small finite clusters and a general slowing down of dynamics due to the large $\Delta$ in conjunction with the large connectivity of the triangular lattice.

\subsection{Generation of kink free initial states}

\begin{figure}
    \centering
    \includegraphics{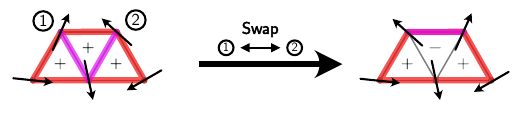}
    \caption{Kink removal Monte-Carlo update used to generate initial states for the softcore compression in the domain wall kink free model. The sign of chirality on each triangle is indicated by $+/-$. Edges where the hardcore constraint is marginally fulfilled at $\Delta = 2\pi/5$ are indicated in red; domain walls of chirality are indicated in pink.}
    \label{fig:kink_removal_move}
\end{figure}

We generate kink-free initial states by starting with a random state and subsequently removing kinks by a combination of a special ``kink-removing'' Monte-Carlo update and regular single-spin-randomization dynamics. The ``kink-removing'' Monte-Carlo update is shown in \autoref{fig:kink_removal_move}. It consists of a swap of the two spins on the shorter of the two parallel edges of the parallelogram defined by the three adjacent triangles of equal chirality. Performing multiple sweeps of such updates on a random state at $\Delta = 0$ removes most, but not all domain wall kinks. We therefore alternate such sweeps with regular single-spin-randomization Monte-Carlo sweeps, but with the additional constraint that no single update should increase the total number of kinks in the system. This procedure after a few iterations successfully removes all domain wall kinks from the state, without inducing any obvious form of chiral or orientational order.  

\subsection{Distribution of $\Delta_J$}

\begin{figure}
    \centering
    \includegraphics{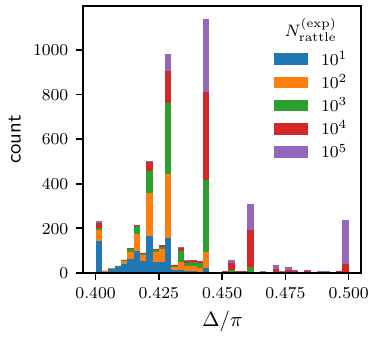}
    \caption{Histogram of exclusion angles $\Delta$ at which the softcore protocol (\autoref{alg:softcore-compression}) for the kink free model jams, for different $N^{\rm (exp)}_{\rm rattle}$. For $N_{\rm rattle}^{(\rm exp)}= 10^5$, 189 of the 1000 runs reach the 120-degree state at $\Delta = 2\pi/3$. }
    \label{fig:kink_free_delta_J}
\end{figure}

Using the kink-free initial states at $\Delta = 0$, we now run the softcore compression protocol (\autoref{alg:softcore-compression}) with the additional constraint on the dynamics that no update is allowed to introduce a domain wall kink into the system. 

In \autoref{fig:kink_free_delta_J}, we show a histogram of the exclusion angles $\Delta_J$ at which the protocol terminates for $10^3$ initial states. 
First, in contrast to the model with kinks (see \autoref{fig:hists}) here we find a broad distribution of $\Delta_J$. We attribute this somewhat surprising result to the fact that for  $\Delta > 2\pi/5$, the system finds a multitude of ways to exhibit slow dynamics with contributions from both, the nonuniversal physics of small clusters as well as the general slowing down at these exclusion angles. Note that both of these phenomena are more important on the triangular lattice than on the square lattice because of the larger connectivity of the former. The pronounced peaks on top of the broad distributions are located at rational fractions of $2\pi$ and are a consequence of what we call \emph{Ouroboros} configurations. These are configurations on a closed loop $\mathcal C$ of length $L$ on the lattice, such that 
\begin{equation}
    \phi_j = \frac{2\pi n j}{L},~~j\in \mathcal C
\end{equation}
for some integer $n$. They are local minima of the energy functional in Eq. (4) of the main text and become forbidden at exclusion angles $\Delta = 2\pi n/L$. At these values of $\Delta$, they introduce long but finite relaxation timescales into the system (see also discussion in the main text).
These attributions are evidenced by the fact that as $N_{\rm rattle}^{(\rm exp)}$ is increased, that is compression is performed more slowly, the broad, continuous distribution gives more and more away to a collection of peaks.
Second, the weight of the distribution shifts to larger $\Delta_J$ as $N_{\rm rattle}^{(\rm exp)}$ is increased and for $N_{\rm rattle}^{(\rm exp)}= 10^5$, 189 of the 1000 runs even reach the 120-degree state at $\Delta = 2\pi/3$. However, no run ever fails between $\pi/2$ and $2\pi/3$.
This implies that smooth domain walls of chirality are indeed the last defects to become forbidden by the hardcore constraint.

\bibliography{references}